\documentclass[aip,amsmath,amssymb,
preprint]{revtex4-1}

\usepackage{graphicx}
\usepackage{dcolumn}
\usepackage{stackengine}
\usepackage{bm}
\usepackage[utf8]{inputenc}
\usepackage[T1]{fontenc}
\usepackage{mathptmx}
\usepackage{array}
\renewcommand{\refname}{}

\begin{document}
\title[Characterization of Folding and 
Stretching in Mixing 
Enhancements]{Characterization of Folding 
and Stretching in Mixing Enhancements}

\author{J. Rodríguez}
 \email{jarodriguez12@uc.cl}

\affiliation{ 
Departamento de Ingeniería Mecánica y 
Metalúrgica, Pontificia Universidad 
Católica de Chile, 7820244, Santiago, 
Chile.
}

\affiliation{ 
Micro and Nanofluidics Laboratory for Life 
Sciences, Pontificia Universidad Católica 
de Chile, 7820244, Santiago, Chile.}

\author{D. Chen}%

\affiliation{ 
Departamento de Ingeniería Mecánica, 
Universidad de Santiago de Chile, 9170020, 
Santiago, Chile.}

\affiliation{ 
Micro and Nanofluidics Laboratory for Life 
Sciences, Pontificia Universidad Católica 
de Chile, 7820244, Santiago, Chile.}

\author{A. M. Guzmán}

\affiliation{Departamento de Ingeniería 
Industrial, Universidad Diego Portales, 
Santiago, Chile.}

\affiliation{ 
Solar and Thermal Energy Conversion and 
Storage Device and System Laboratory, 
Santiago, Chile}

\begin{abstract}
This research paper presents a 2D 
numerical model of an electrokinetic 
T-junction micro-mixer based on the 
stretching and folding theory presented by 
Ottino. Particle deformation was considered
by simulating 2 $\mu m$ massless particles 
as 8-point square cells. Furthermore, 
stretching and folding definitions are 
proposed, compatible with a Lagrangian 
particle approach. Moreover, mixing 
homogeneity and consistency were measured 
in a 200 $\mu m$ square region of interest 
neighboring the outlet. Statistical 
analysis of the exiting mixing homogeneity 
at four different electric field conditions
(93.5 $V/cm$, 109.8 $V/cm$, 126 $V/cm$ and 
117.9 $V/cm$, corresponding to a 23$V$, 
27$V$, 29$V$ and 31V potential difference) 
show that mixing consistency and 
homogeneity are not always increased with a
higher electric field intensity, even after
electrokinetic instabilities are formed, as
increasingly unstable flow conditions 
decrease the ratio of folding to stretching
($m$), hindering the interaction between 
substances. Finally, an optimal proportion of stretching to folding was found for 
maximizing mixing efficiency at $m 
=0.0045$. 
\end{abstract}
\maketitle

\section{\label{sec:level1} Introduction} 
Over the past 20 years there has been 
extensive research going into reducing the 
time and cost for chemical analysis and 
pathogen detection. Particularly, 
development of chip devices for micro total
analysis systems ($\mu$TAS) has been on the
rise. These systems are often composed of a
series of microchannels to operate, and as 
such, the study of small-scale channel 
flows has led the way for various areas of study; one of them being 
electrokinetic (EK) flow. EK flow is a type
of flow which is driven by an electric 
field \cite{Gill1990}. In terms of physical
variables, this field studies the coupling 
of fluid flow, electric fields and the 
transport of species. In the past, because 
of the scale of these $\mu$TAS  apparatus, 
achieving high Reynolds numbers by using 
pressure gradients, to induce turbulence 
and catalyze chemical reactions, proved to 
be a challenge, as mixing of species would 
occur primarily by diffusion. Diffusion is 
not a hasty process, and as such, mixing 
channel sizes were constrained to be 
larger. As a result, several designs were 
created to ameliorate the mixing of species
inside by modifying the geometries 
\cite{Wang2010, Stroock2002,Bakker2000}, 
however, thanks to the study of EK flows, 
it was soon proved that turbulence can be 
induced in laminar flows \cite{Wang2014}, 
in the form of electrokinetic instabilities
(EKI), allowing for smaller and more 
efficient micromixers. The study of EKI 
formation and the determination of 
parameters, such as the electrical Rayleigh
number $Ra_e$ \cite{Chen2005, 
Baygents1998,Lin2004} to ensure 
reproducibility of this newfound turbulence
would pave the way for new sophisticated 
micromixer setups. From pulsating electric 
field ideas \cite{Glasgow2004,Kang2008} to 
diverse geometries 
\cite{Azimi2017,Posner2006, Li2016} 
including herringbone passive mixers 
\cite{Stroock2002, Camesasca2006, 
Aubin2003}, several steps were taken to 
improve mixture homogeneity. Theory wise, 
important advances on chaotic mixing 
theory\cite{Baker1990, Ottino1989} were 
made, particularly describing the mixing 
process as a series of transformations that
the fluid undergoes by stretching and 
folding until a homogenous product is 
achieved.

In the present, significant 
contributions have been made on the further
study of stretching and its mixing 
enhancements \cite{Kelley2011,Voth2002, 
Arratia2005}, yet, there is still a lack of 
consensus on how both of these deformation 
processes (stretching and folding) impact 
directly, the mixing homogeneity in EK 
flows. In this investigation, a 2D 
T-junction geometry is studied via 
Finite Element Method (FEM) simulations and
a relationship between folding, stretching 
and mixing quality is found. Moreover, a 
validated model is proposed, compatible 
with a Lagrangian particle approach. Mixing
consistency through time is measured using 
statistical methods, by analyzing a region 
of interest neighboring the outlet of the 
microchannel. Finally, voltage is related 
to stretching and folding and their rate of
change in the short term.

\section{Methods and materials}
\subsection{Folding and Stretching}

In this work we investigate the role of 
stretching and folding and their influence 
in flow mixing. A series of 2D simulations 
were carried out using the FEM, to study 
their effect in a T - shaped microchannel. 
Based on Ottino’s \cite{Ottino1989} theory 
of stretching and folding, mixing was 
characterized by the reduction of 
curvilinear fluid filament length scales. 
Adapting this general proposition to a 
Lagrangian, particle simulation, gives rise
to practical definitions for studying both 
mechanisms in a two-dimensional domain.

\begin{figure}[h!]
    \centering
    \includegraphics[scale = 0.4]{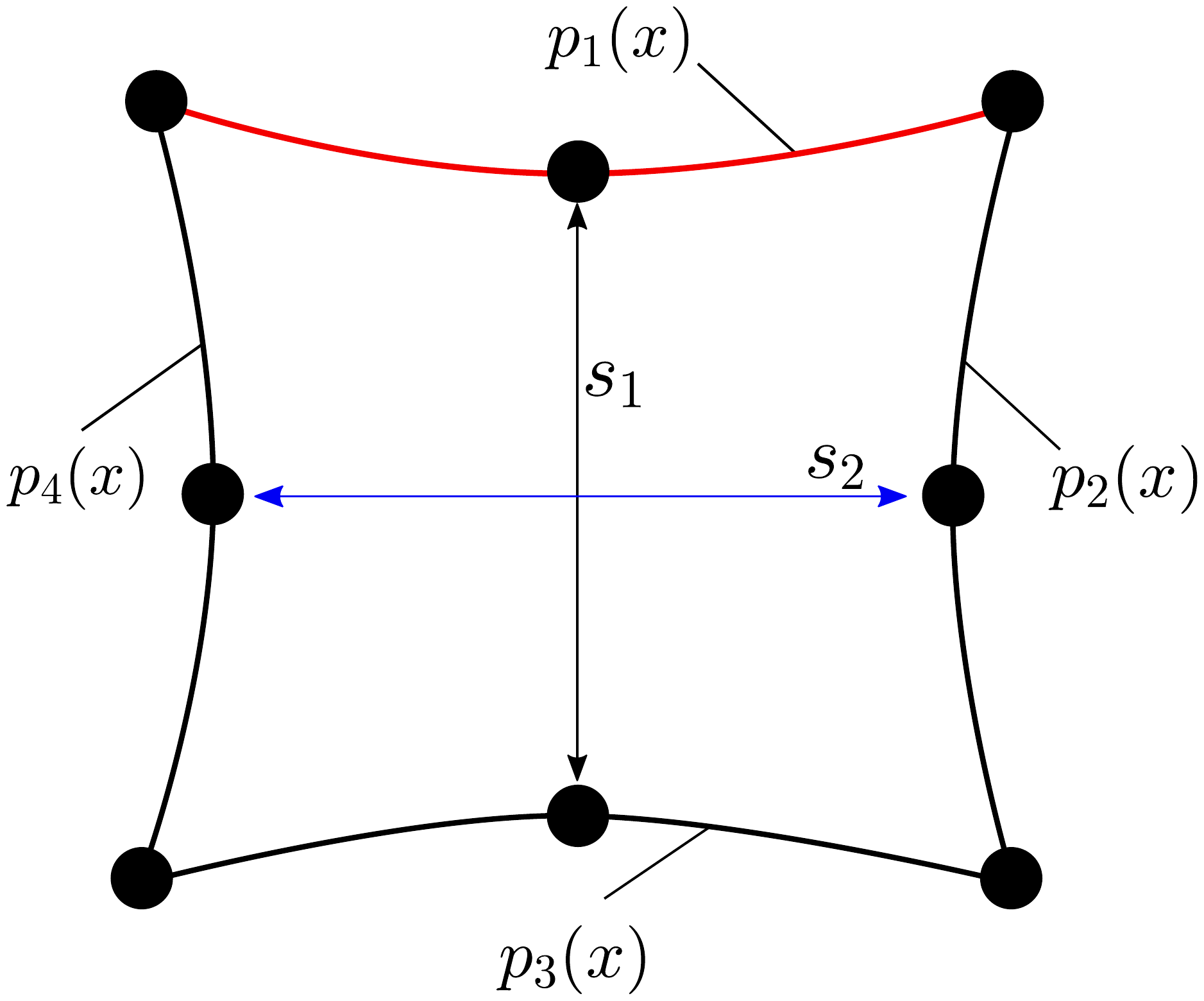}
    \caption{Diagram of eight-point massless cell subjected to stretching and folding.}
    \label{fig:folding}
\end{figure}

Originally, stretching ($\lambda$) was 
defined by Ottino \cite{Ottino1989} as the 
deformation of an infinitesimal fluid 
length, i.e.,

\begin{equation}
    \lambda = \lim_{|d \mathbf{X}|\to 0} \frac{d \mathbf{x}}{d\mathbf{X}}.
\end{equation}
Where, $|d\mathbf{X}|$ is the magnitude of 
the initial stretch vector at initial time 
$t_0=0$ and $|d\mathbf{x}|$ is the 
magnitude of the stretch vector at time 
$t$. Adapting this definition to a discrete particle approach, it follows that,

\begin{equation}
    S_{1j}=\left(\frac{\Delta s_1}{s_1}\right)_j , 
\end{equation}

\begin{equation}
    S_{2j}=\left(\frac{\Delta s_2}{s_2}\right)_j , 
\end{equation}
where $S_{1_j}$ and $S_{2_j}$ represent the stretching of the initial lengths $s_1$ and
$s_2$ shown on Fig. \ref{fig:folding}. The 
subscript $j$ is referent to the cell $j$. 

Figure 1 shows the massless cell, composed 
by eight points, used to evaluate 
stretching and folding. Where $P_i (x), (i 
= 1, 2, 3, 4)$ are the best-fit, second 
order polynomial functions that join the 
points on each of the four sides of the 
cell and x is the streamwise direction.

The principal stretching direction out of 
$s_1$ and $s_2$, for each cell is defined 
using Eq.~(\ref{eq:four}) as,

\begin{equation}
    S_j = \max(S_{1j},S_{2j})
    \label{eq:four}.
\end{equation}
The stretching average value for both 
directions for all massless cells analyzed,
is calculated by Eq.~(\ref{eq:five}) as,

\begin{equation}
    \bar{S} = \frac{1}{N} \sum_{j}^N S_j
    \label{eq:five}.
\end{equation}
where $N$ is the total number of analyzed 
cells, at a given time.

Analogously, the folding can be interpreted
as the bending deformation of the cell 
$\varepsilon$, derived from the Euler - 
Bernoulli beam theory, shown by Hibbeler, 
(2010) \cite{Hibbeler2010} in 
Eq.~(\ref{eq:six}).

\begin{equation}
    \varepsilon = \frac{-y}{R} =-\kappa y, 
    \label{eq:six}
\end{equation}
where $R$ is the curvature radius, $y$ is 
the distance from the neutral axis and 
$\kappa$ is the curvature. When the 
distance y is shorter than  2 $\mu m$, 
Eq.~(\ref{eq:six}) can be simplified and 
rewritten as Eq.~(\ref{eq:seven}).

\begin{equation}
    \varepsilon = \pm \frac{1 \mu m}{R} =\pm \kappa y. 
    \label{eq:seven}
\end{equation}
Since the direction of the bending or 
folding of the cell is irrelevant for the 
purposes of fluid mixing, we propose a new 
general definition which must be applicable
to all discrete cells in the simulated 
flow. Based on  the mathematical definition
of curvature for  one-dimensional functions
\cite{Stewart2010}, the curvature for each 
side of the cell can be expressed by Eq.~(\ref{eq:eight}).

\begin{equation}
    k_i = \frac{|p_i '' (x)|}{\left(1+p_i'(x)^2\right)^\frac{3}{2}} \Bigg\rvert_{x=x_2}
    \label{eq:eight}
\end{equation}
where $p_i'' (x)$ is the second derivative 
and $p_i'$ (x) is the first derivative of 
$p_i (x)$ with respect to the variable $x$,
respectively. Moreover, $x_2$ is the 
mid-point of the side $i$. Given $k_i$, which
corresponds to the curvature of one side of
the cell, then the maximum curvature of the
cell can be determined using 
Eq.~(\ref{eq:nine}). 

\begin{equation}
    K_j = \max(k_1, k_2, k_3, k_4)
    \label{eq:nine}
\end{equation}
Where the subscript $j$ refers to the cell that is being analyzed at a given time. The complete folding process is characterized by averaging the $K_j$ values over N cells at a given time given by Eq.~(\ref{eq:ten}):

\begin{equation}
    \bar{F} = \frac{1}{N} \sum_{j}^N K_j
    \label{eq:ten}
\end{equation}
The total average deformation, that considers the folding (bending) and stretching effects is defined as the sum of both, see Eq.~(\ref{eq:eleven}). 

\begin{equation}
    \nu  = \bar{S}+\bar{F}
    \label{eq:eleven}
\end{equation}

\subsection{Mixing index}

To correlate the quality of the mixer to the stretching and folding. The following mixing index expression, previously defined by other authors \cite{Bakker2000,Kang2008,Wu2003,Lu2001,Jayaraj2007} will be employed to evaluate quantitatively the homogeneity of the mixture. 

\begin{equation}
    D = \sqrt{\frac{1}{N} \sum_{k,l}^N \left( 1-\frac{c_{k,l}}{\bar{c}}\right)}
\end{equation}
Where $c_{k,l}$ is the concentration of the node $(k,l)$, $\bar{c}$ is the spatial average of concentration for the whole fluid domain, and $N$ is the total number of data points in the sample. $D$ then indicates the coefficient of variation, CoV, of the spatial concentration. As $D$ decreases, the mixture concentration deviates less from the mean, becoming more homogeneous.

Defining mixing efficiency, as a quantity that increases proportionally with mixture homogeneity, is more intuitive and therefore, defined by Eq.~(\ref{eq:thirteen}) as,

\begin{equation}
    \text{Mixing efficiency} = 100 \cdot (1-D)
    \label{eq:thirteen}
\end{equation}

\medskip

The focus of this paper is based on the mixture leaving the microchannel, hence, a 200 $\mu m$ box neighboring the outlet, pictured in Figure 2, was designated as the interest region to study the D value. This way, the whole width of the microchannel is considered while the effects of upstream turbulence are dampened, so that only EKI affect the $D$ values.

\subsection{Numerical Simulation}

A 2D electrokinetic flow was simulated in a
T-junction microchannel, using the finite 
element method (FEM). A time-dependent 
approach was used, for varying simulated 
time spans,  considering the simulation 
time is double than the characteristic time
$t^*=\frac{|\mathbf{V}|}{L_T}$  
\cite{Kang2006}¸ with $L_T$ the  distance 
from the inlet to the outlet of the 
microchannel and $\mathbf{V}$ the flow 
average velocity. In order to effectively 
create chaotic motion, a series of 
simulations with a range of increasing 
electric potential $\phi$  were carried out
to find the most suitable conditions for 
flow instabilities (EKI). Additionally, all
simulations were carried out for constant 
values of the conductivity ratio 
$\gamma=\frac{\sigma_H}{\sigma_L}$  and 
zeta potential $\zeta$. 
Table~\ref{tab:table1} shows the fluid 
properties considered in our study.

\begin{table}[h!]
\caption{\label{tab:table1}Fluid Properties used for the numerical simulation.}
\centering
\begin{tabular}{>{\centering\arraybackslash}p{2cm}>{\centering\arraybackslash}p{2cm}>{\centering\arraybackslash}p{2cm}>{\centering\arraybackslash}p{2cm}>{\centering\arraybackslash}p{3cm}>{\centering\arraybackslash}p{2cm}>{\centering\arraybackslash}p{2cm}}
\hline
$\rho$        & $D_{eff}$     & $\mu$        & $\zeta$     & $\varepsilon$                             & $\sigma_H$ & $\sigma_L$ \\ \hline
1000 $\frac{kg}{m3}$ & 1.989$\cdot 10^9 \frac{m^2}{s}$ & 0.001 $Pa \cdot s$ & -0.088 V & 6.933 $\cdot 10^{-10} \frac{F}{m}$ & 0.15 $\frac{S}{m}$  & 0.01 $\frac{S}{m}$  \\ \hline
\end{tabular}
\label{table:1}
\end{table}

In this analysis, the governing equations 
of mass conservation for an incompressible 
and Newtonian fluid
(Eq.~(\ref{eq:14})), momentum balance 
(including the electrical driving force) 
(Eq.~(\ref{eq:15})), electric current 
conservation (Eq.~(\ref{eq:16})) and 
species conservation (Eq.~(\ref{eq:17})) 
are solved simultaneously.

\begin{equation}
    \nabla \cdot \mathbf{V} = 0
    \label{eq:14}
\end{equation}

\begin{equation}
    \rho \left(\frac{\partial \mathbf{V}}{\partial t} + \mathbf{V} \cdot (\nabla \mathbf{V}) \right) = 
    -\nabla p+\mu \nabla ^2 \mathbf{V} +\rho_f \mathbf{E}
    \label{eq:15}
\end{equation}

\begin{equation}
    \frac{\partial \rho_e}{\partial t}+\nabla \cdot \mathbf{J} = 0
    \label{eq:16}
\end{equation}

\begin{equation}
    \frac{\partial C}{\partial t} + \mathbf{V} \cdot (\nabla C) = \nabla \cdot (D_{eff} \nabla C)
    \label{eq:17}
\end{equation}

For Eqs.~(\ref{eq:14}) and (\ref{eq:15}), $\mathbf{V}$
is the velocity field, $p$ is the pressure 
field, $\rho$ is the density of the fluid, 
$\mu$ is the dynamic viscosity of the fluid
and $\mathbf{E}$ is the electric field. The
charge density $\rho_f$  is defined by 
Gauss’ Law \cite{Gill1990} as $\rho_f=\varepsilon  
\nabla \cdot \mathbf{E}$, where 
$\mathbf{E}$ itself is defined by 
$\mathbf{E}=-\nabla \phi$ and the 
permittivity is defined as 
$\varepsilon=\varepsilon_r \varepsilon_0$. 
Where $\varepsilon_r$ and $\varepsilon_0$ 
are the dielectric constant for water and 
universal permittivity, respectively. In Eq.~\ref{eq:16},  $\rho_e$ is the electric current 
density, $\mathbf{J}$ is the total 
electrical current of the system, $C$ is 
the species concentration and $D_{\text{eff}}$ is 
the effective diffusivity, given by 
\cite{Posner2006, Luo2009}:

\begin{equation}
    D_{\text{eff}} = \frac{2D_+ +D_-}{D_+ +D_-} .
    \label{eq:19}
\end{equation}
Where $D_{\pm}$ is the cationic/anionic 
diffusivity coefficient, defined as 
$D_{\pm}=RTm_{\pm}$. Parameter $R$ is the 
universal gas constant, $T$ is the absolute 
temperature and $m_{\pm}$ is the ionic 
mobility.

In order to solve the PDE system of Eqs.~(\ref{eq:14}) – (\ref{eq:17}), a highly non-linear 
Newton-Raphson solver was used, with a 
minimum dampening factor of 0.0001. To 
characterize the EKI the electrical 
Rayleigh number $Ra_e=\frac{\varepsilon E^2
\frac{d}{2}}{\mu D}$ defined by Hao et al. 
(2004) \cite{Lin2004} was used. Where EKI 
form only after a certain critical 
electrical Rayleigh number $Ra_{e_{cr}}$ has 
been reached. Once the velocity field was 
determined, the particle trajectories were 
calculated by solving the kinematics 
$\frac{d \vec{q}}{dt}=\mathbf{V}$ by the 
Lagrangian approach, using a constant Newton-Raphson method.

\begin{figure}[h!]
    \centering
    \includegraphics[width=0.7\textwidth]{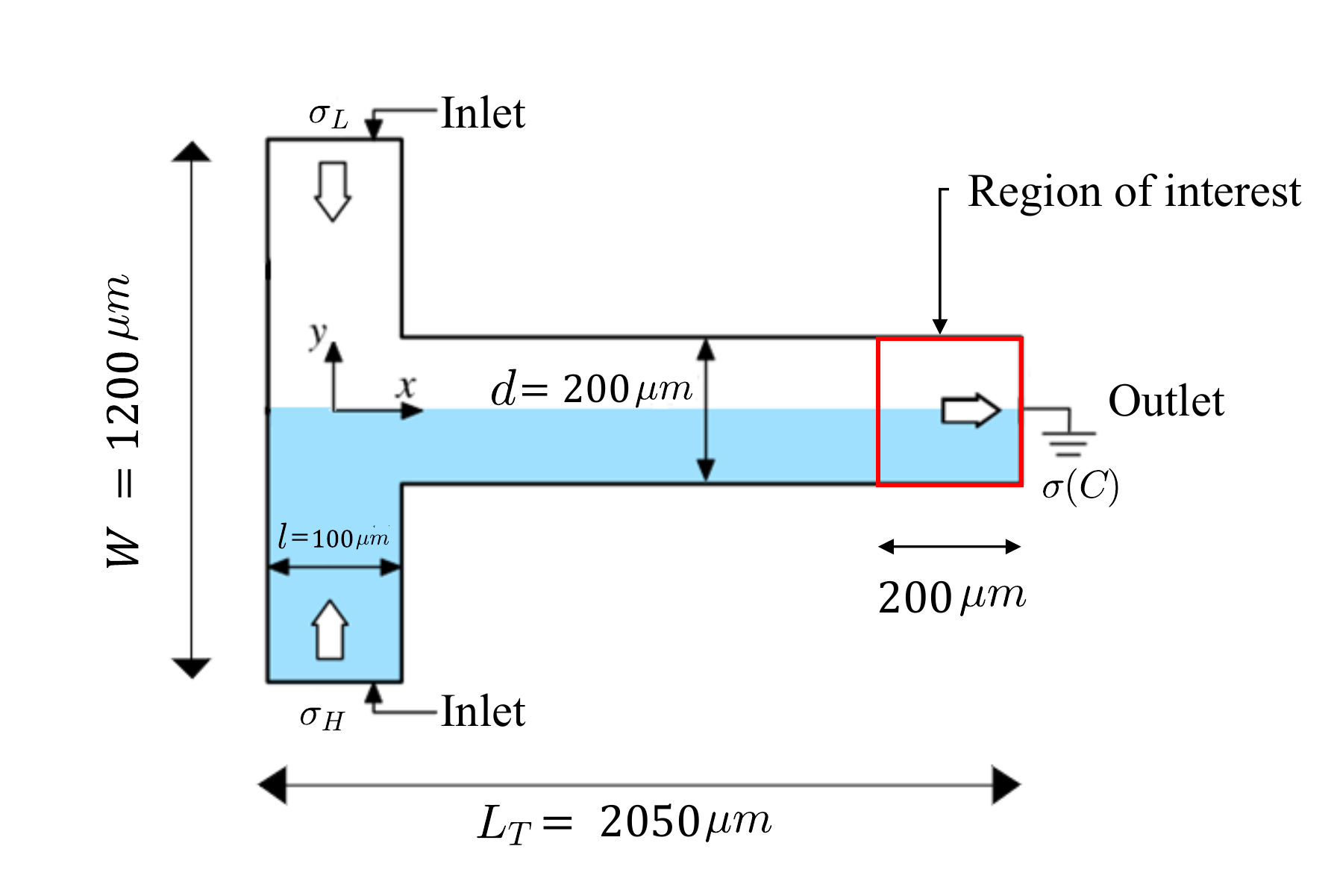}
    \caption{2D Computational domain for the numerical model, where the blue area represents a high conductivity fluid and the white area represents a low conductivity one. The red box encloses the domain where mixing will be investigated.}
    \label{fig:setup}
\end{figure}

Figure \ref{fig:setup} shows the T-junction computational
domain that was simulated, along with its dimensions. These include: the distance 
between the upper and lower inlets  $W=1200$ 
$\mu m$, the microchannel outlet and inlet 
widths $d=200 \mu m$ and $l=100 \mu m$, respectively, and the microchannel total 
length $L_T=2050 \mu m$. The lower inlet 
allowed flow of high electric conductivity 
$\sigma_H$ fluid while the upper inlet 
allowed low electric conductivity 
$\sigma_L$ fluid into the domain. We 
propose Eq.~\ref{eq:19} in order to take into 
account the temporal and spatial variation
of the electrical conductivity $\sigma$ 
which may occur during the mixing process 
of two miscible fluids with different 
concentrations $C$.

\begin{equation}
\sigma(C)= \sigma_H \bigg(\frac{C}{C_{ref}} \bigg) + \sigma_L \bigg( 1-\frac{C}{C_{ref}}\bigg),
\end{equation}
\begin{equation*}
\centering
C_{ref} \equiv 1; \begin{cases} 
      \text{if}  &\text{$C=1 \rightarrow \sigma =\sigma_H$} \\
      \text{if}  &\text{ $C=1 \rightarrow \sigma =\sigma_L$} \\
      if  &\text{ $0<C<1 \rightarrow \sigma_H < \sigma < \sigma_L$}
\end{cases}
\label{eq:19}
\end{equation*}

The following boundary conditions were 
implemented. For species conservation, a 
constant concentration $C= 0 
\frac{mol}{m^3}$  and $C= 1 
\frac{mol}{m^3}$  was set up in the upper 
and lower inlet respectively. For the 
momentum conservation equation, a value of 0 $Pa$ was considered for
the inlet and outlet pressure, 
respectively. Finally, for the current 
conservation equation, the outlet voltage condition was 
$\phi_{out}=0\text{V}$ and for both inlets equal voltage values were set $\phi_{in_1 
}=\phi_{in_2}$ for each of the cases studied, which
varied inlet voltage from 23V to 31V. Additionally, 
the wall boundary conditions considered 
were: no flow, electroosmotic slip velocity
and electrically isolated for the three 
conservation laws respectively.

To generate the deformable massless cells in
space, we released a grid arrangement of 96
particles from each inlet. Each 
deformable cell was composed by eight 
particles, in order to mimic a 2 $\mu m$ 
diameter massless cell as shown in Figure \ref{fig:folding}. The particle conditions used in the 
Particle Tracing module include: the 
particle type, which was set to liquid and 
massless, the particle dynamic viscosity 
$\mu_p$ and particle surface tension 
$\sigma_p$ values, which were 1 mPa $\cdot$
s and 0.0729 N/m respectively. The particle
charge value was Z = 0, neglecting the charge 
interactions.

\section{Results and Discussion}

\subsection{Model Validation}

The numerical model’s EKI behavior was 
proved identical to the literature, as 
shown in Figure 3.

\begin{figure}[h!]
    \includegraphics[width=\textwidth]{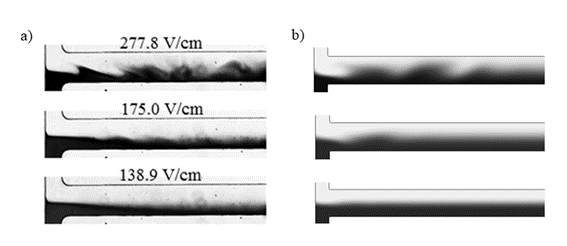}
    \caption{Interfacial behavior comparison for: (a) (0.2 $\times$ EMG 408) and water co-flow in a 45 $\mu m$ deep T-shaped microchannel. High conductivity ferrofluid in black and water in white. Experimental data from Song et al. 2017 \cite{Song2017}.  (b) 109.8 V/cm simulated flow. High conductivity solution in black and low conductivity solution in white.}
    \label{fig:comparison}
\end{figure}

Proving similarity to this particular set 
of experimental results by Song et al. 2017 \cite{Song2017} is significant, as the 
dimensions of the setup closely match those
used in this study. The compared flows were
observed on a T-junction setup with 
dimensions of length  $L_T$=2000 $\mu m$, 
distance between the upper and lower inlets
$W=700$ $\mu m$ and outlet and inlet widths $d=200$ $\mu m$ and $l=100$ $\mu m$.

Figure 4 shows the mixing efficiency 
obtained in 29V flow during $t=5.2s$, 
compared to the normalized results reported
by Luo et al. 2008 \cite{Winjet2008}. There
is a close resemblance in trends, as the 
average value for the simulated results is 
of 62\% mixing efficiency whereas Luo et 
al. 2008’s results show 50\% mixing 
efficiency. As both position and time 
affect the mixing efficiency, finding a 
similarity in behavior from two different 
flows, at different times is substantial, 
as the geometry analyzed by Luo et al. 2008
is comparable to the computational domain 
used for simulations. The T-junction setup 
used in the literature had dimensions of 
length  $L_T$=1005 $\mu m$, distance 
between the upper and lower inlets $W=708$ 
$\mu m$ and outlet and inlet widths $d=60$ 
$\mu m$ and $l=60$ $\mu m$, respectively.

\begin{figure}[h!]
    \centering
    \includegraphics{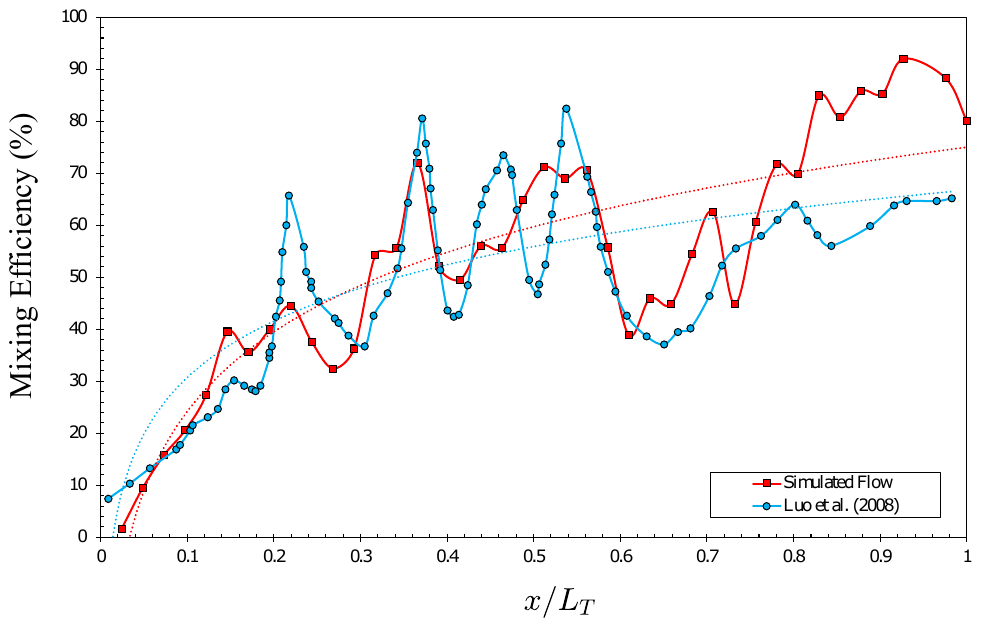}
    \caption{Evolution of mixing efficiency through dimensionless space. 40 datapoints from $x= 0$ $\mu m$ to $x= 2050$ $\mu m$ are shown in red for a 29V simulated flow at $t=5.2s$. Normalized results from Luo et al. 2008 \cite{Winjet2008}, for a 750 V/cm flow at $t=0.3s$ are shown in blue.}
    \label{fig:mixingcomp}
\end{figure}

\subsection{Importance of total deformation in mixing efficiency}

\begin{figure}[h]
    \centering
    \includegraphics[scale=0.82]{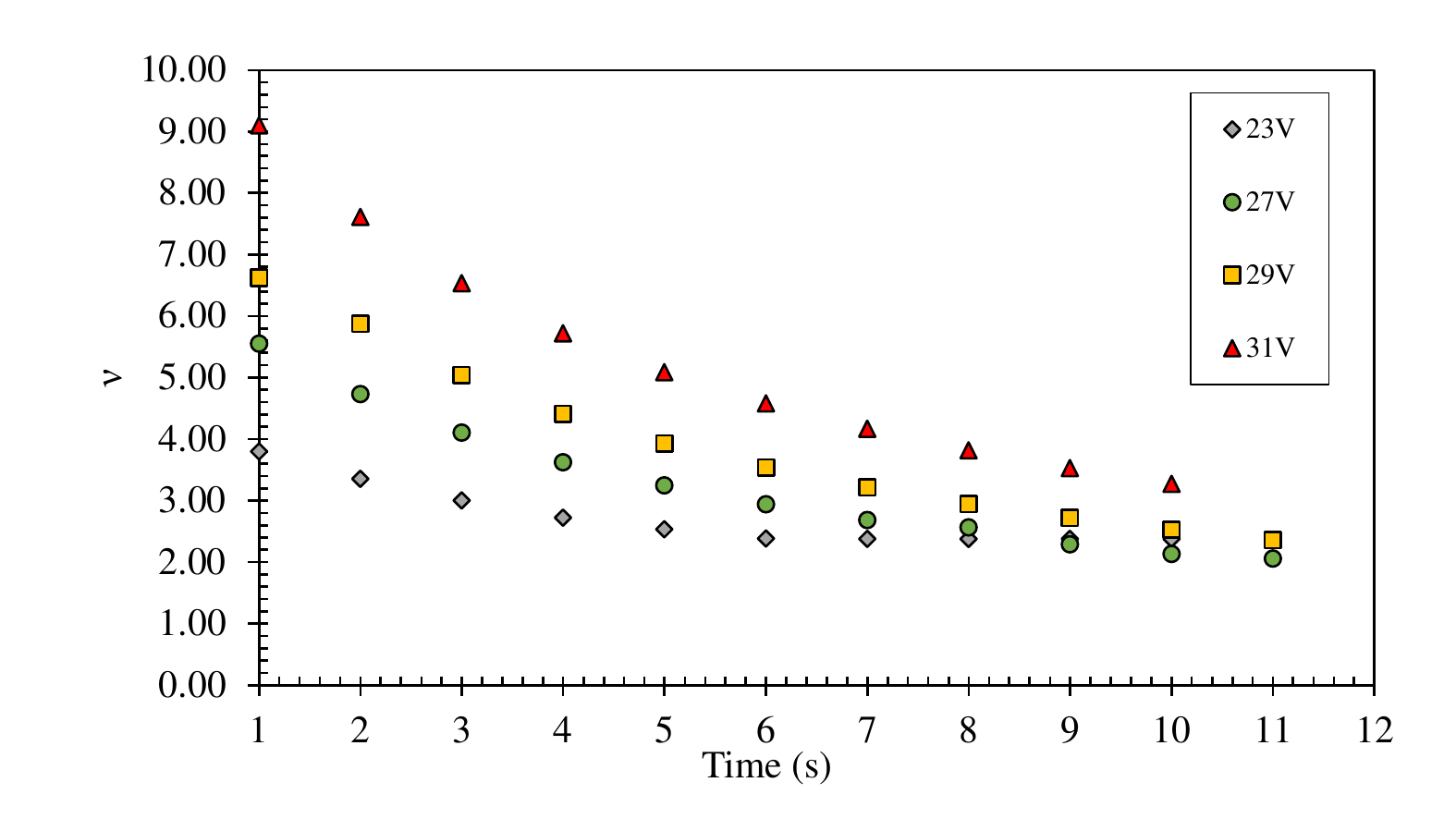}
    \caption{Evolution of $\nu$ with time for 23V, 27V, 29V and 31V flows. 11 timesteps were simulated. Each timestep size was set to 1s.}
    \label{fig:nu}
\end{figure}

Figure 5 shows the change in total 
deformation ($\nu$) with time, for four 
different voltage conditions. As voltage 
increases, $\nu$ increases as well, 
generating a steeper curve that flattens as
voltage is decreased. For all flows, it is 
evident that regardless of the voltage, 
$\nu$  decreases with time. For lower 
voltage flows, however, the rate of change 
of the total deformation is smaller than 
for the higher voltage flows. Horizontality
of the results can be measured by the total
deformation change $\Delta \nu$ in the 
timeframe analyzed. A higher $\Delta \nu$ 
value represents less horizontality. For 
instance, for the 31V flow a $\Delta \nu$ 
value of 5.83 was calculated, the 29V flow 
showed a $\Delta \nu$ value of 4.26 and the
27V flow’s $\Delta \nu$ was measured at  
3.50, whereas the 23V flow, in the same 
timeframe, had a change in its total 
deformation of 1.48; plateauing earlier 
than the other flows, after 6 timesteps. As the voltage across the 
geometry decreases, EKI dampen, $\nu (t)$ 
becomes more horizontal and the flow 
stabilizes in a smaller time window. 

To determine if statistically there is an 
impact in the consistency of the mixture at
higher voltages, 50 datapoints were 
extracted for each of the flows that showed
EKI.

\begin{table}[h!]
\caption{\label{tab:table2}D Value Intervals for 23V, 27V, 29V and 31V flows.}
\resizebox{\textwidth}{!}{%
\begin{tabular}{>{\centering\arraybackslash}p{2.4cm}>{\centering\arraybackslash}p{2.4cm}>{\centering\arraybackslash}p{2.4cm}>{\centering\arraybackslash}p{2.4cm}>{\centering\arraybackslash}p{2.4cm}>{\centering\arraybackslash}p{2.4cm}>{\centering\arraybackslash}p{2.4cm}}
\hline
\textbf{Voltage}     & \multicolumn{6}{c}{\textbf{Intervals}}                                                                                                                                                                                                                                                                                                                                                                                                  \\ \hline
\multicolumn{1}{l}{} & \textbf{\begin{tabular}
[c]{@{}c@{}}1\\ {[}$a=0 < b${]}
\end{tabular}} & \textbf{\begin{tabular}[c]{@{}c@{}}2\\ {[}$a - b${]}\end{tabular}} & \textbf{\begin{tabular}[c]{@{}c@{}}3\\ {[}$a - b${]}\end{tabular}} & \textbf{\begin{tabular}[c]{@{}c@{}}4\\ {[}$a - b${]}\end{tabular}} & \textbf{\begin{tabular}[c]{@{}c@{}}5\\ {[}$a - b${]}\end{tabular}} & \textbf{\begin{tabular}[c]{@{}c@{}}6\\ {[}$a >b=0${]}\end{tabular}} \\
23V                  & \textless 0.291                                                            & 0.291 - 0.320                                                    & 0.320 - 0.349                                                    & 0.349 - 0.379                                                    & 0.379 - 0.408                                                    & \textgreater 0.408                                                             \\
27V                  & \textless 0.123                                                            & 0.123 - 0.198                                                    & 0.198 - 0.273                                                    & 0.273 - 0.349                                                    & 0.349 - 0.424                                                    & \textgreater 0.424                                                             \\
29V                  & \textless 0.000*                                                           & 0.000 - 0.112*                                                   & 0.112 - 0.231                                                    & 0.231 - 0.350                                                    & 0.350 - 0.469                                                    & \textgreater 0.469                                                             \\ 
31V                  & \textless 0.063                                                            & 0.063 - 0.151                                                    & 0.151 - 0.238                                                    & 0.238 - 0.326                                                    & 0.326 - 0.413                                                    & \textgreater 0.413                                                             \\ \hline
\end{tabular}%
}
\end{table}

Table~\ref{tab:table2} shows the intervals chosen for the statistical analysis shown in Figure 6, where $a$ is the lower bound and $b$ is the upper bound of each interval, respectively. Intervals 1 - 6 were evenly spaced by one standard deviation. Note that in the case of the 29V flow, the last two intervals reach the minimum value for $D$. This happens because the average $D$ value of the 50 datapoints analyzed for this flow was very small in relation to its standard deviation, and as so, the last intervals fell out of range.  

Figure 6 shows four histograms relating the
frequency of the mixing value $D$ for the 
intervals previously defined. From first to
sixth, these intervals go from the lowest $D$
value (most homogeneous mixture) to 
highest. Each graph is presented with its 
normal trend in red. As the potential 
difference across the channel increases, 
the $D$ value distribution begins to shift to
the left, decreasing deviation from the 
average concentration and thus, improving 
the mixing efficiency. This is evidenced in
(a) by the consistent, yet, heterogeneous 
mixture of the stable 23V flow, as $D$ was 
found in the range of [0.349 - 0.379] on 72\%
of the cases. As EKI are generated in (b), 
the 27V flow shows an increase in mixture 
homogeneity and a decrease in consistency, 
as the sample is more evenly distributed 
along the intervals. As the flow continues 
to grow unstable, the 29V flow in (c) 
reaches peak consistency, concentrating 
54\% of the sample in the third interval 
and 62\% of it in the second and third 
intervals. As the first three intervals now
hold the highest frequency, the homogeneity
of the mixture has been improved, producing
a consistent $D$ value in the range of 
[0.000 - 0.231]. The 31V flow in (d), closely
follows the consistency and homogeneity 
obtained in (c) by concentrating 58\% of 
the sample on the first three intervals, 
corresponding to the $D$ value range of 
[<0.0631 - 0.238].

\begin{figure}[ht]
    \centering
    \includegraphics[width = \textwidth]{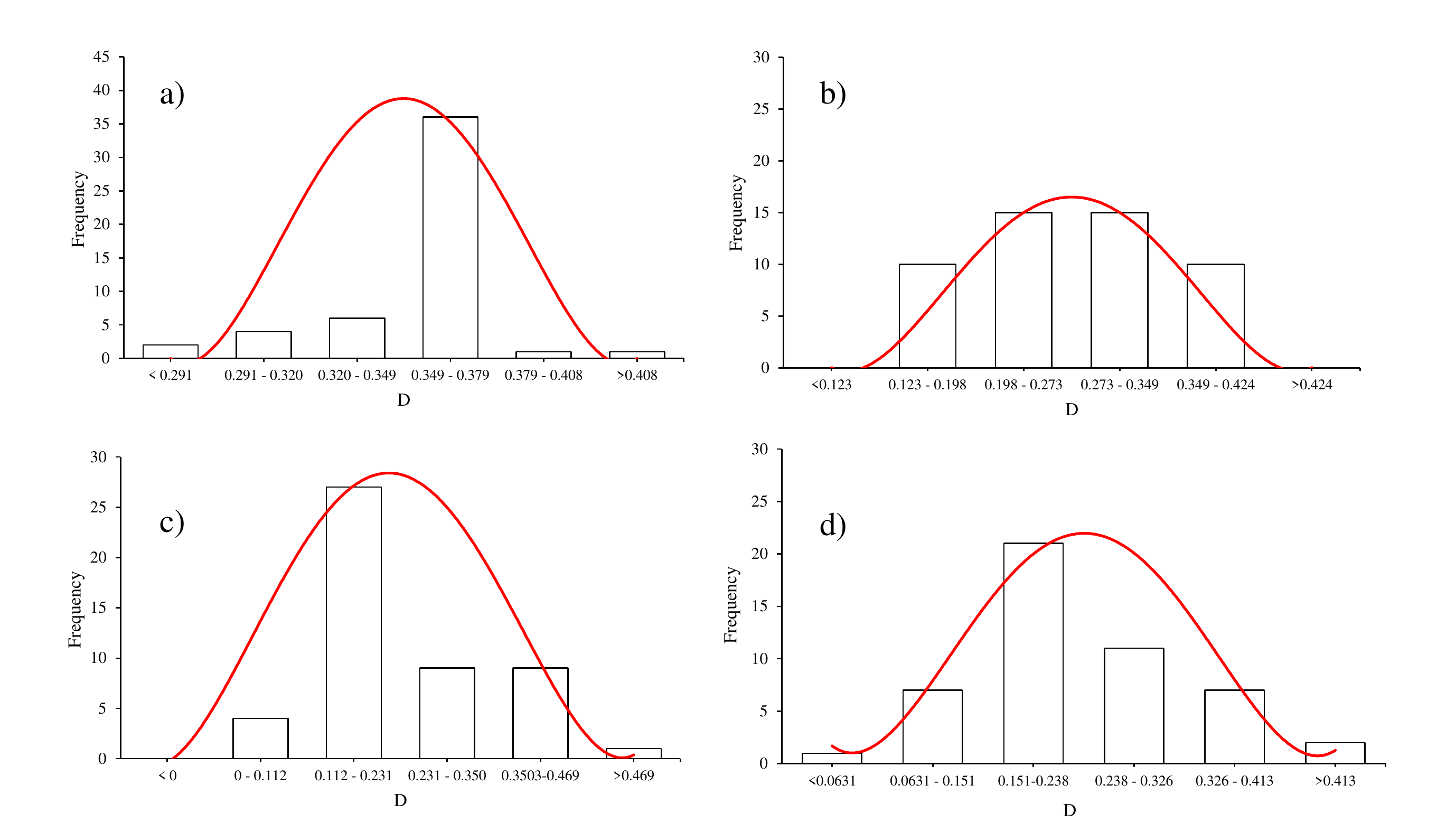}
    \caption{Effect of potential difference on mixing efficiency frequency distribution for (a) 23V where no EKI were formed, (b) 27V EKI present, (c) 29V and (d) 31V. 50 different datapoints were taken for each histogram. The timestep difference between each datapoint was 0.2s.}
    \label{fig:six}
\end{figure}

As higher voltages are studied $\nu$ 
increases as reported in Figure \ref{fig:nu}. However,
the 31V displayed the highest value for 
$\nu$, but it did not have the most 
consistent mixture through time. This shows
that further increasing $\nu$ in the 
microchannel, after 29V does not improve 
mixing consistency. 

\begin{figure}[h]
    \centering
    \includegraphics[scale=0.8]{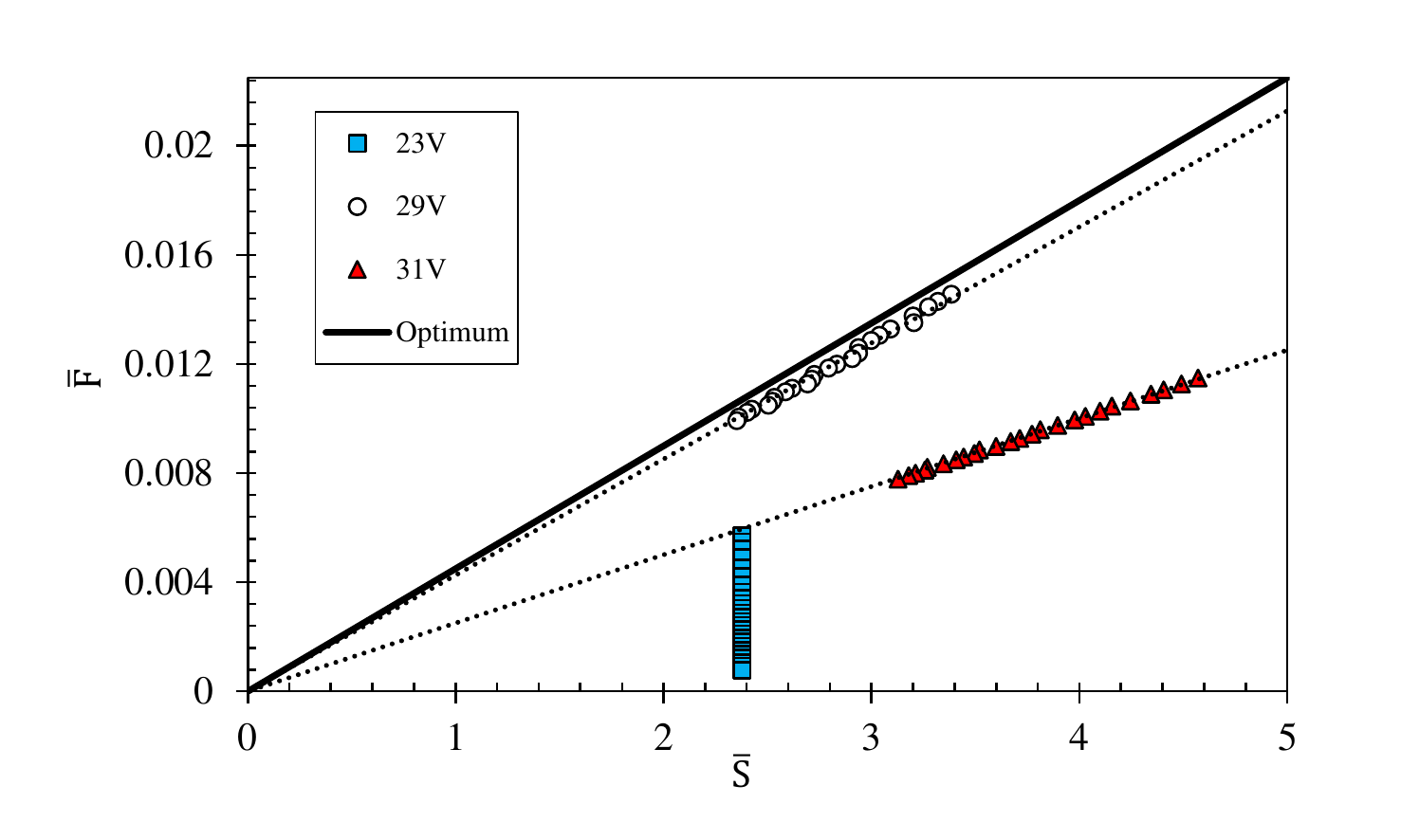}
    \caption{Effect of potential difference on proportion between $\bar{F}$ and $\bar{S}$. 23V (square), 29V (dot) and 31V (triangle) flows were analyzed. 25 datapoints were studied per flow. Each datapoint represents a different time and the timestep was set to 0.2s.}
    \label{fig:seven}
\end{figure}

Figure~\ref{fig:seven} shows the relationship between 
$\bar{S}$ and $\bar{F}$ for three different 
flow conditions. As established in Figure~\ref{fig:nu}, all flows stabilize with time in terms 
of $\nu$, and the same holds true in Figure~\ref{fig:seven}, where as time passes, the datapoints 
decrease in terms of both $\bar{S}$ and 
$\bar{F}$. To ensure fully developed flow, 
out of the total simulated time of 11 
seconds, only the last 5 seconds were 
plotted for this analysis. Moreover, to 
precisely measure the proportional change 
of $\bar{F}$ in relation to $\bar{S}$, the 
slope $m$ was defined by Eq.~\ref{eq:20},

\begin{equation}
    m=\frac{\bar{F}}{\bar{S}}
    \label{eq:20}
\end{equation}

A stable 23V flow is depicted in Figure~\ref{fig:seven}, 
which plateaus at a stretching value 
$\bar{S}$ of 2.376 while showing a 
pronounced proportion of folding to 
stretching of $m\to \infty$. As time 
passes, only folding increases in a stable 
flow, and therefore $\nu$ varies 
insignificantly.

After EKI are reached at 
27V, further increasing the voltage to 29V 
displaces the datapoints to another 
trendline, with a slope $m$ of 0.0043. When
comparing the trend of the 31V results to 
the ones obtained from the 29V flow, it can
be observed that the main difference in 
particle deformation is the proportion of 
folding to stretching $m$, as the 31V flow 
always holds a higher value of  $\nu$  for 
every time period analyzed (established in 
Figure~\ref{fig:nu}). Realizing that the slope of the 
most unstable flow (31V) is the one that 
most closely resembles the least unstable 
flow (23V) at  $m= 0.0025$, explains why in
Figure~\ref{fig:six} the highest voltage flow did not 
achieve the most consistent mixture. As 
fluid velocity increases with the electric 
field, and the latter increases with 
potential difference, a faster flow might 
stretch the particles more, yet, it lowers 
the chances of particle folding to occur, 
as full bends have less time to develop. As
the 29V slope appears to be oriented 
halfway between the vertical 23V slope and 
the more horizontal 31V results, an upper 
boundary symbolizing the optimum mixing 
conditions for the T-section studied is 
proposed in Figure 7. A completely 
proportional $\bar{F}$ to $\bar{S}$ 
relationship was found at $m=0.0045$. When 
this condition is reached, the maximum 
mixing efficiency is obtained, as the ratio
of folding to stretching is at an optimum, 
where cells have just enough time to 
perform full bends while also being 
stretched by the velocity of the flow.

\section{Conclusion}

A numerical study of a T-junction geometry 
is presented in this paper. Particle 
kinematics were considered in the 
simulation, by releasing massless points in
arrangements of 8-point cells. By 
characterizing and measuring the stretching
and folding processes undergone by these 
cells, it was determined that further 
increasing the electric field in EK 
microflows does not necessarily improve the
mixing quality of the exiting substance. 
Furthermore, mixing consistency in the 
short term was related to the proportion of
folding to stretching m by comparing the 
evolution of the total deformation for the 
23V, 27V, 29V and 31V flows for 11s, to the
frequency distribution of the mixing values
recorded. After EKI are reached at 27V, 
further increasing the electric field only 
improves mixing (both consistency in the 
short term and homogeneity) if $m$ increases. 
A higher sample concentration, of 62\% is 
found for the first three intervals of $D$
values for the 29V flow when compared to 
the 31V flow that concentrates only 58\%. $D$
values lower than 0.2 were reached only 
after 20 seconds of simulation time, 
achieving a more homogeneous mixture than 
Wu et al.’s (2003) \cite{Wu2003} 
micromixer, which obtained a minimum 0.8 $D$ 
value on their constant $\zeta$ potential 
value results, and assuring an equally 
homogeneous mixture as Kang et al. (2008) 
\cite{Kang2008} in a significantly smaller 
time frame (less than 30 timesteps). 
Moreover, the flow that produces the most 
homogeneous mixture and highest mixing 
consistency (29V) shows a higher $m$ value of
0.0043 than the second most consistent and 
homogeneous mixing flow (31V), which holds 
a value of 0.0025. An optimal ratio of 
folding to stretching was determined, as an
upper bound to the mixing efficiency at 
$m=0.0045$.

\begin{acknowledgements}
The authors are thankful for the computing resources lent by Jaime Middleton in this investigation.

\end{acknowledgements}

\section{References}
\bibliography{references}

\begin{thebibliography}{29}%
\makeatletter
\providecommand \@ifxundefined [1]{%
 \@ifx{#1\undefined}
}%
\providecommand \@ifnum [1]{%
 \ifnum #1\expandafter \@firstoftwo
 \else \expandafter \@secondoftwo
 \fi
}%
\providecommand \@ifx [1]{%
 \ifx #1\expandafter \@firstoftwo
 \else \expandafter \@secondoftwo
 \fi
}%
\providecommand \natexlab [1]{#1}%
\providecommand \enquote  [1]{``#1''}%
\providecommand \bibnamefont  [1]{#1}%
\providecommand \bibfnamefont [1]{#1}%
\providecommand \citenamefont [1]{#1}%
\providecommand \href@noop [0]{\@secondoftwo}%
\providecommand \href [0]{\begingroup \@sanitize@url \@href}%
\providecommand \@href[1]{\@@startlink{#1}\@@href}%
\providecommand \@@href[1]{\endgroup#1\@@endlink}%
\providecommand \@sanitize@url [0]{\catcode `\\12\catcode `\$12\catcode
  `\&12\catcode `\#12\catcode `\^12\catcode `\_12\catcode `\%12\relax}%
\providecommand \@@startlink[1]{}%
\providecommand \@@endlink[0]{}%
\providecommand \url  [0]{\begingroup\@sanitize@url \@url }%
\providecommand \@url [1]{\endgroup\@href {#1}{\urlprefix }}%
\providecommand \urlprefix  [0]{URL }%
\providecommand \Eprint [0]{\href }%
\providecommand \doibase [0]{http://dx.doi.org/}%
\providecommand \selectlanguage [0]{\@gobble}%
\providecommand \bibinfo  [0]{\@secondoftwo}%
\providecommand \bibfield  [0]{\@secondoftwo}%
\providecommand \translation [1]{[#1]}%
\providecommand \BibitemOpen [0]{}%
\providecommand \bibitemStop [0]{}%
\providecommand \bibitemNoStop [0]{.\EOS\space}%
\providecommand \EOS [0]{\spacefactor3000\relax}%
\providecommand \BibitemShut  [1]{\csname bibitem#1\endcsname}%
\let\auto@bib@innerbib\@empty
\bibitem [{\citenamefont {Gill}(1990)}]{Gill1990}%
  \BibitemOpen
  \bibfield  {author} {\bibinfo {author} {\bibfnamefont {W.}~\bibnamefont
  {Gill}},\ }\bibfield  {title} {\enquote {\bibinfo {title} {{Physicochemical
  Hydrodynamics, an Introduction}},}\ }\href {\doibase
  10.1016/0301-9322(90)90046-l} {\bibfield  {journal} {\bibinfo  {journal}
  {International Journal of Multiphase Flow}\ }\textbf {\bibinfo {volume}
  {16}},\ \bibinfo {pages} {167--168} (\bibinfo {year} {1990})}\BibitemShut
  {NoStop}%
\bibitem [{\citenamefont {Wang}\ and\ \citenamefont {Hu}(2010)}]{Wang2010}%
  \BibitemOpen
  \bibfield  {author} {\bibinfo {author} {\bibfnamefont {C.~T.}\ \bibnamefont
  {Wang}}\ and\ \bibinfo {author} {\bibfnamefont {Y.~C.}\ \bibnamefont {Hu}},\
  }\bibfield  {title} {\enquote {\bibinfo {title} {{Mixing of liquids using
  obstacles in Y-type microchannels}},}\ }\href {\doibase
  10.6180/jase.2010.13.4.04} {\bibfield  {journal} {\bibinfo  {journal}
  {Tamkang Journal of Science and Engineering}\ }\textbf {\bibinfo {volume}
  {13}},\ \bibinfo {pages} {385--394} (\bibinfo {year} {2010})}\BibitemShut
  {NoStop}%
\bibitem [{\citenamefont {Stroock}\ \emph {et~al.}(2002)\citenamefont
  {Stroock}, \citenamefont {Dertinger}, \citenamefont {Ajdari}, \citenamefont
  {Mezi{\'{c}}}, \citenamefont {Stone},\ and\ \citenamefont
  {Whitesides}}]{Stroock2002}%
  \BibitemOpen
  \bibfield  {author} {\bibinfo {author} {\bibfnamefont {A.~D.}\ \bibnamefont
  {Stroock}}, \bibinfo {author} {\bibfnamefont {S.~K.}\ \bibnamefont
  {Dertinger}}, \bibinfo {author} {\bibfnamefont {A.}~\bibnamefont {Ajdari}},
  \bibinfo {author} {\bibfnamefont {I.}~\bibnamefont {Mezi{\'{c}}}}, \bibinfo
  {author} {\bibfnamefont {H.~A.}\ \bibnamefont {Stone}}, \ and\ \bibinfo
  {author} {\bibfnamefont {G.~M.}\ \bibnamefont {Whitesides}},\ }\bibfield
  {title} {\enquote {\bibinfo {title} {{Chaotic mixer for microchannels}},}\
  }\href {\doibase 10.1126/science.1066238} {\bibfield  {journal} {\bibinfo
  {journal} {Science}\ }\textbf {\bibinfo {volume} {295}},\ \bibinfo {pages}
  {647--651} (\bibinfo {year} {2002})}\BibitemShut {NoStop}%
\bibitem [{\citenamefont {Bakker}, \citenamefont {Laroche},\ and\ \citenamefont
  {Marshall}(2000)}]{Bakker2000}%
  \BibitemOpen
  \bibfield  {author} {\bibinfo {author} {\bibfnamefont {A.}~\bibnamefont
  {Bakker}}, \bibinfo {author} {\bibfnamefont {R.~D.}\ \bibnamefont {Laroche}},
  \ and\ \bibinfo {author} {\bibfnamefont {E.~M.}\ \bibnamefont {Marshall}},\
  }\bibfield  {title} {\enquote {\bibinfo {title} {{Laminar Flow in Static
  Mixers with Helical Elements}},}\ }\href@noop {} {\bibfield  {journal}
  {\bibinfo  {journal} {The Online CFM book}\ ,\ \bibinfo {pages} {1 -- 11}}
  (\bibinfo {year} {2000})}\BibitemShut {NoStop}%
\bibitem [{\citenamefont {Wang}, \citenamefont {Yang},\ and\ \citenamefont
  {Zhao}(2014)}]{Wang2014}%
  \BibitemOpen
  \bibfield  {author} {\bibinfo {author} {\bibfnamefont {G.~R.}\ \bibnamefont
  {Wang}}, \bibinfo {author} {\bibfnamefont {F.}~\bibnamefont {Yang}}, \ and\
  \bibinfo {author} {\bibfnamefont {W.}~\bibnamefont {Zhao}},\ }\bibfield
  {title} {\enquote {\bibinfo {title} {{There can be turbulence in
  microfluidics at low Reynolds number}},}\ }\href {\doibase
  10.1039/c3lc51403j} {\bibfield  {journal} {\bibinfo  {journal} {Lab on a
  Chip}\ }\textbf {\bibinfo {volume} {14}},\ \bibinfo {pages} {1452--1458}
  (\bibinfo {year} {2014})}\BibitemShut {NoStop}%
\bibitem [{\citenamefont {Chen}\ \emph {et~al.}(2005)\citenamefont {Chen},
  \citenamefont {Lin}, \citenamefont {Lele},\ and\ \citenamefont
  {Santiago}}]{Chen2005}%
  \BibitemOpen
  \bibfield  {author} {\bibinfo {author} {\bibfnamefont {C.~H.}\ \bibnamefont
  {Chen}}, \bibinfo {author} {\bibfnamefont {H.}~\bibnamefont {Lin}}, \bibinfo
  {author} {\bibfnamefont {S.~K.}\ \bibnamefont {Lele}}, \ and\ \bibinfo
  {author} {\bibfnamefont {J.~G.}\ \bibnamefont {Santiago}},\ }\bibfield
  {title} {\enquote {\bibinfo {title} {{Convective and absolute electrokinetic
  instability with conductivity gradients}},}\ }\href {\doibase
  10.1017/S0022112004002381} {\bibfield  {journal} {\bibinfo  {journal}
  {Journal of Fluid Mechanics}\ }\textbf {\bibinfo {volume} {524}},\ \bibinfo
  {pages} {263--303} (\bibinfo {year} {2005})}\BibitemShut {NoStop}%
\bibitem [{\citenamefont {Baygents}\ and\ \citenamefont
  {Baldessari}(1998)}]{Baygents1998}%
  \BibitemOpen
  \bibfield  {author} {\bibinfo {author} {\bibfnamefont {J.~C.}\ \bibnamefont
  {Baygents}}\ and\ \bibinfo {author} {\bibfnamefont {F.}~\bibnamefont
  {Baldessari}},\ }\bibfield  {title} {\enquote {\bibinfo {title}
  {{Electrohydrodynamic instability in a thin fluid layer with an electrical
  conductivity gradient}},}\ }\href {\doibase 10.1063/1.869567} {\bibfield
  {journal} {\bibinfo  {journal} {Physics of Fluids}\ }\textbf {\bibinfo
  {volume} {10}},\ \bibinfo {pages} {301--311} (\bibinfo {year}
  {1998})}\BibitemShut {NoStop}%
\bibitem [{\citenamefont {Lin}\ \emph {et~al.}(2004)\citenamefont {Lin},
  \citenamefont {Storey}, \citenamefont {Oddy}, \citenamefont {Chen},\ and\
  \citenamefont {Santiago}}]{Lin2004}%
  \BibitemOpen
  \bibfield  {author} {\bibinfo {author} {\bibfnamefont {H.}~\bibnamefont
  {Lin}}, \bibinfo {author} {\bibfnamefont {B.~D.}\ \bibnamefont {Storey}},
  \bibinfo {author} {\bibfnamefont {M.~H.}\ \bibnamefont {Oddy}}, \bibinfo
  {author} {\bibfnamefont {C.~H.}\ \bibnamefont {Chen}}, \ and\ \bibinfo
  {author} {\bibfnamefont {J.~G.}\ \bibnamefont {Santiago}},\ }\bibfield
  {title} {\enquote {\bibinfo {title} {{Instability of electrokinetic
  microchannel flows with conductivity gradients}},}\ }\href {\doibase
  10.1063/1.1710898} {\bibfield  {journal} {\bibinfo  {journal} {Physics of
  Fluids}\ }\textbf {\bibinfo {volume} {16}},\ \bibinfo {pages} {1922--1935}
  (\bibinfo {year} {2004})}\BibitemShut {NoStop}%
\bibitem [{\citenamefont {Glasgow}, \citenamefont {Batton},\ and\ \citenamefont
  {Aubry}(2004)}]{Glasgow2004}%
  \BibitemOpen
  \bibfield  {author} {\bibinfo {author} {\bibfnamefont {I.}~\bibnamefont
  {Glasgow}}, \bibinfo {author} {\bibfnamefont {J.}~\bibnamefont {Batton}}, \
  and\ \bibinfo {author} {\bibfnamefont {N.}~\bibnamefont {Aubry}},\ }\bibfield
   {title} {\enquote {\bibinfo {title} {{Electroosmotic mixing in
  microchannels}},}\ }\href {\doibase 10.1039/b408875a} {\bibfield  {journal}
  {\bibinfo  {journal} {Lab on a Chip}\ }\textbf {\bibinfo {volume} {4}},\
  \bibinfo {pages} {558--562} (\bibinfo {year} {2004})}\BibitemShut {NoStop}%
\bibitem [{\citenamefont {Kang}, \citenamefont {Heo},\ and\ \citenamefont
  {Suh}(2008)}]{Kang2008}%
  \BibitemOpen
  \bibfield  {author} {\bibinfo {author} {\bibfnamefont {J.}~\bibnamefont
  {Kang}}, \bibinfo {author} {\bibfnamefont {H.~S.}\ \bibnamefont {Heo}}, \
  and\ \bibinfo {author} {\bibfnamefont {Y.~K.}\ \bibnamefont {Suh}},\
  }\bibfield  {title} {\enquote {\bibinfo {title} {{LBM simulation on mixing
  enhancement by the effect of heterogeneous zeta-potential in a
  microchannel}},}\ }\href {\doibase 10.1007/s12206-008-0301-4} {\bibfield
  {journal} {\bibinfo  {journal} {Journal of Mechanical Science and
  Technology}\ }\textbf {\bibinfo {volume} {22}},\ \bibinfo {pages}
  {1181--1191} (\bibinfo {year} {2008})}\BibitemShut {NoStop}%
\bibitem [{\citenamefont {Azimi}, \citenamefont {Nazari},\ and\ \citenamefont
  {Daghighi}(2017)}]{Azimi2017}%
  \BibitemOpen
  \bibfield  {author} {\bibinfo {author} {\bibfnamefont {S.}~\bibnamefont
  {Azimi}}, \bibinfo {author} {\bibfnamefont {M.}~\bibnamefont {Nazari}}, \
  and\ \bibinfo {author} {\bibfnamefont {Y.}~\bibnamefont {Daghighi}},\
  }\bibfield  {title} {\enquote {\bibinfo {title} {{Developing a fast and
  tunable micro-mixer using induced vortices around a conductive flexible
  link}},}\ }\href {\doibase 10.1063/1.4975982} {\bibfield  {journal} {\bibinfo
   {journal} {Physics of Fluids}\ }\textbf {\bibinfo {volume} {29}} (\bibinfo
  {year} {2017}),\ 10.1063/1.4975982}\BibitemShut {NoStop}%
\bibitem [{\citenamefont {Posner}\ and\ \citenamefont
  {Santiago}(2006)}]{Posner2006}%
  \BibitemOpen
  \bibfield  {author} {\bibinfo {author} {\bibfnamefont {J.~D.}\ \bibnamefont
  {Posner}}\ and\ \bibinfo {author} {\bibfnamefont {J.~G.}\ \bibnamefont
  {Santiago}},\ }\bibfield  {title} {\enquote {\bibinfo {title} {{Convective
  instability of electrokinetic flows in a cross-shaped microchannel}},}\
  }\href {\doibase 10.1017/S0022112005008542} {\bibfield  {journal} {\bibinfo
  {journal} {Journal of Fluid Mechanics}\ }\textbf {\bibinfo {volume} {555}},\
  \bibinfo {pages} {1--42} (\bibinfo {year} {2006})}\BibitemShut {NoStop}%
\bibitem [{\citenamefont {Li}, \citenamefont {Delorme},\ and\ \citenamefont
  {Frankel}(2016)}]{Li2016}%
  \BibitemOpen
  \bibfield  {author} {\bibinfo {author} {\bibfnamefont {Q.}~\bibnamefont
  {Li}}, \bibinfo {author} {\bibfnamefont {Y.}~\bibnamefont {Delorme}}, \ and\
  \bibinfo {author} {\bibfnamefont {S.~H.}\ \bibnamefont {Frankel}},\
  }\bibfield  {title} {\enquote {\bibinfo {title} {{Parametric numerical study
  of electrokinetic instability in cross-shaped microchannels}},}\ }\href
  {\doibase 10.1007/s10404-015-1666-1} {\bibfield  {journal} {\bibinfo
  {journal} {Microfluidics and Nanofluidics}\ }\textbf {\bibinfo {volume}
  {20}},\ \bibinfo {pages} {1--12} (\bibinfo {year} {2016})}\BibitemShut
  {NoStop}%
\bibitem [{\citenamefont {Camesasca}, \citenamefont {Kaufman},\ and\
  \citenamefont {Manas-Zloczower}(2006)}]{Camesasca2006}%
  \BibitemOpen
  \bibfield  {author} {\bibinfo {author} {\bibfnamefont {M.}~\bibnamefont
  {Camesasca}}, \bibinfo {author} {\bibfnamefont {M.}~\bibnamefont {Kaufman}},
  \ and\ \bibinfo {author} {\bibfnamefont {I.}~\bibnamefont
  {Manas-Zloczower}},\ }\bibfield  {title} {\enquote {\bibinfo {title}
  {{Staggered passive micromixers with fractal surface patterning}},}\ }\href
  {\doibase 10.1088/0960-1317/16/11/008} {\bibfield  {journal} {\bibinfo
  {journal} {Journal of Micromechanics and Microengineering}\ }\textbf
  {\bibinfo {volume} {16}},\ \bibinfo {pages} {2298--2311} (\bibinfo {year}
  {2006})}\BibitemShut {NoStop}%
\bibitem [{\citenamefont {Aubin}\ \emph {et~al.}(2003)\citenamefont {Aubin},
  \citenamefont {Fletcher}, \citenamefont {Bertrand},\ and\ \citenamefont
  {Xuereb}}]{Aubin2003}%
  \BibitemOpen
  \bibfield  {author} {\bibinfo {author} {\bibfnamefont {J.}~\bibnamefont
  {Aubin}}, \bibinfo {author} {\bibfnamefont {D.~F.}\ \bibnamefont {Fletcher}},
  \bibinfo {author} {\bibfnamefont {J.}~\bibnamefont {Bertrand}}, \ and\
  \bibinfo {author} {\bibfnamefont {C.}~\bibnamefont {Xuereb}},\ }\bibfield
  {title} {\enquote {\bibinfo {title} {{Characterization of the mixing quality
  in micromixers}},}\ }\href {\doibase 10.1002/ceat.200301848} {\bibfield
  {journal} {\bibinfo  {journal} {Chemical Engineering and Technology}\
  }\textbf {\bibinfo {volume} {26}},\ \bibinfo {pages} {1262--1270} (\bibinfo
  {year} {2003})}\BibitemShut {NoStop}%
\bibitem [{\citenamefont {Baker}, \citenamefont {Gollub},\ and\ \citenamefont
  {Fox}(1990)}]{Baker1990}%
  \BibitemOpen
  \bibfield  {author} {\bibinfo {author} {\bibfnamefont {G.~L.}\ \bibnamefont
  {Baker}}, \bibinfo {author} {\bibfnamefont {J.~P.}\ \bibnamefont {Gollub}}, \
  and\ \bibinfo {author} {\bibfnamefont {R.}~\bibnamefont {Fox}},\ }\bibfield
  {title} {\enquote {\bibinfo {title} {{ Chaotic Dynamics: An Introduction
  }},}\ }\href {\doibase 10.1063/1.2810630} {\bibfield  {journal} {\bibinfo
  {journal} {Physics Today}\ } (\bibinfo {year} {1990}),\
  10.1063/1.2810630}\BibitemShut {NoStop}%
\bibitem [{\citenamefont {Ottino}(1989)}]{Ottino1989}%
  \BibitemOpen
  \bibfield  {author} {\bibinfo {author} {\bibfnamefont {J.~M.}\ \bibnamefont
  {Ottino}},\ }\bibfield  {title} {\enquote {\bibinfo {title} {{The kinematics
  of mixing: stretching, chaos, and transport}},}\ }\href@noop {} {\  (\bibinfo
  {year} {1989})}\BibitemShut {NoStop}%
\bibitem [{\citenamefont {Kelley}\ and\ \citenamefont
  {Ouellette}(2011)}]{Kelley2011}%
  \BibitemOpen
  \bibfield  {author} {\bibinfo {author} {\bibfnamefont {D.~H.}\ \bibnamefont
  {Kelley}}\ and\ \bibinfo {author} {\bibfnamefont {N.~T.}\ \bibnamefont
  {Ouellette}},\ }\bibfield  {title} {\enquote {\bibinfo {title} {{Separating
  stretching from folding in fluid mixing}},}\ }\href {\doibase
  10.1038/nphys1941} {\bibfield  {journal} {\bibinfo  {journal} {Nature
  Physics}\ }\textbf {\bibinfo {volume} {7}},\ \bibinfo {pages} {477--480}
  (\bibinfo {year} {2011})}\BibitemShut {NoStop}%
\bibitem [{\citenamefont {Voth}, \citenamefont {Haller},\ and\ \citenamefont
  {Gollub}(2002)}]{Voth2002}%
  \BibitemOpen
  \bibfield  {author} {\bibinfo {author} {\bibfnamefont {G.~A.}\ \bibnamefont
  {Voth}}, \bibinfo {author} {\bibfnamefont {G.}~\bibnamefont {Haller}}, \ and\
  \bibinfo {author} {\bibfnamefont {J.~P.}\ \bibnamefont {Gollub}},\ }\bibfield
   {title} {\enquote {\bibinfo {title} {{Experimental Measurements of
  Stretching Fields in Fluid Mixing}},}\ }\href {\doibase
  10.1103/PhysRevLett.88.254501} {\bibfield  {journal} {\bibinfo  {journal}
  {Physical Review Letters}\ }\textbf {\bibinfo {volume} {88}},\ \bibinfo
  {pages} {4} (\bibinfo {year} {2002})}\BibitemShut {NoStop}%
\bibitem [{\citenamefont {Arratia}, \citenamefont {Voth},\ and\ \citenamefont
  {Gollub}(2005)}]{Arratia2005}%
  \BibitemOpen
  \bibfield  {author} {\bibinfo {author} {\bibfnamefont {P.~E.}\ \bibnamefont
  {Arratia}}, \bibinfo {author} {\bibfnamefont {G.~A.}\ \bibnamefont {Voth}}, \
  and\ \bibinfo {author} {\bibfnamefont {J.~P.}\ \bibnamefont {Gollub}},\
  }\bibfield  {title} {\enquote {\bibinfo {title} {{Stretching and mixing of
  non-Newtonian fluids in time-periodic flows}},}\ }\href {\doibase
  10.1063/1.1909184} {\bibfield  {journal} {\bibinfo  {journal} {Physics of
  Fluids}\ }\textbf {\bibinfo {volume} {17}},\ \bibinfo {pages} {1--10}
  (\bibinfo {year} {2005})}\BibitemShut {NoStop}%
\bibitem [{\citenamefont {Hibbeler}(2010)}]{Hibbeler2010}%
  \BibitemOpen
  \bibfield  {author} {\bibinfo {author} {\bibfnamefont {R.~C.}\ \bibnamefont
  {Hibbeler}},\ }\href@noop {} {\emph {\bibinfo {title} {{Mechanics of
  Materials}}}},\ \bibinfo {edition} {8th}\ ed.\ (\bibinfo  {publisher}
  {Prentice Hall},\ \bibinfo {year} {2010})\ pp.\ \bibinfo {pages}
  {255--335}\BibitemShut {NoStop}%
\bibitem [{\citenamefont {Stewart}(2010)}]{Stewart2010}%
  \BibitemOpen
  \bibfield  {author} {\bibinfo {author} {\bibfnamefont {J.}~\bibnamefont
  {Stewart}},\ }\href@noop {} {\emph {\bibinfo {title} {{Calculus: Early
  Trascendentals}}}},\ \bibinfo {edition} {7th}\ ed.\ (\bibinfo  {publisher}
  {Cengage Learning},\ \bibinfo {year} {2010})\ pp.\ \bibinfo {pages}
  {853--860}\BibitemShut {NoStop}%
\bibitem [{\citenamefont {Wu}\ and\ \citenamefont {Liu}(2003)}]{Wu2003}%
  \BibitemOpen
  \bibfield  {author} {\bibinfo {author} {\bibfnamefont {H.~Y.}\ \bibnamefont
  {Wu}}\ and\ \bibinfo {author} {\bibfnamefont {C.~H.}\ \bibnamefont {Liu}},\
  }\bibfield  {title} {\enquote {\bibinfo {title} {{A novel electrokinetic
  micromixer}},}\ }\href {\doibase 10.1109/SENSOR.2003.1215552} {\bibfield
  {journal} {\bibinfo  {journal} {TRANSDUCERS 2003 - 12th International
  Conference on Solid-State Sensors, Actuators and Microsystems, Digest of
  Technical Papers}\ }\textbf {\bibinfo {volume} {1}},\ \bibinfo {pages}
  {631--634} (\bibinfo {year} {2003})}\BibitemShut {NoStop}%
\bibitem [{\citenamefont {Lu}, \citenamefont {Ryu},\ and\ \citenamefont
  {Liu}(2001)}]{Lu2001}%
  \BibitemOpen
  \bibfield  {author} {\bibinfo {author} {\bibfnamefont {L.-H.}\ \bibnamefont
  {Lu}}, \bibinfo {author} {\bibfnamefont {K.~S.}\ \bibnamefont {Ryu}}, \ and\
  \bibinfo {author} {\bibfnamefont {C.}~\bibnamefont {Liu}},\ }\bibfield
  {title} {\enquote {\bibinfo {title} {{A Novel Microstirrer and Arrays for
  Microfluidic Mixing}},}\ }in\ \href {\doibase 10.1007/978-94-010-1015-3_10}
  {\emph {\bibinfo {booktitle} {Micro Total Analysis Systems 2001}}}\ (\bibinfo
  {year} {2001})\BibitemShut {NoStop}%
\bibitem [{\citenamefont {Jayaraj}, \citenamefont {Kang},\ and\ \citenamefont
  {Suh}(2007)}]{Jayaraj2007}%
  \BibitemOpen
  \bibfield  {author} {\bibinfo {author} {\bibfnamefont {S.}~\bibnamefont
  {Jayaraj}}, \bibinfo {author} {\bibfnamefont {S.}~\bibnamefont {Kang}}, \
  and\ \bibinfo {author} {\bibfnamefont {Y.~K.}\ \bibnamefont {Suh}},\
  }\bibfield  {title} {\enquote {\bibinfo {title} {{A review on the analysis
  and experiment of fluid flow and mixing in micro-channels}},}\ }\href
  {\doibase 10.1007/BF02916316} {\bibfield  {journal} {\bibinfo  {journal}
  {Journal of Mechanical Science and Technology}\ }\textbf {\bibinfo {volume}
  {21}},\ \bibinfo {pages} {536--548} (\bibinfo {year} {2007})}\BibitemShut
  {NoStop}%
\bibitem [{\citenamefont {Kang}\ \emph {et~al.}(2006)\citenamefont {Kang},
  \citenamefont {Park}, \citenamefont {Kang},\ and\ \citenamefont
  {Huh}}]{Kang2006}%
  \BibitemOpen
  \bibfield  {author} {\bibinfo {author} {\bibfnamefont {K.~H.}\ \bibnamefont
  {Kang}}, \bibinfo {author} {\bibfnamefont {J.}~\bibnamefont {Park}}, \bibinfo
  {author} {\bibfnamefont {I.~S.}\ \bibnamefont {Kang}}, \ and\ \bibinfo
  {author} {\bibfnamefont {K.~Y.}\ \bibnamefont {Huh}},\ }\bibfield  {title}
  {\enquote {\bibinfo {title} {{Initial growth of electrohydrodynamic
  instability of two-layered miscible fluids in T-shaped microchannels}},}\
  }\href {\doibase 10.1016/j.ijheatmasstransfer.2006.04.026} {\bibfield
  {journal} {\bibinfo  {journal} {International Journal of Heat and Mass
  Transfer}\ }\textbf {\bibinfo {volume} {49}},\ \bibinfo {pages} {4577--4583}
  (\bibinfo {year} {2006})}\BibitemShut {NoStop}%
\bibitem [{\citenamefont {Luo}(2009)}]{Luo2009}%
  \BibitemOpen
  \bibfield  {author} {\bibinfo {author} {\bibfnamefont {W.~J.}\ \bibnamefont
  {Luo}},\ }\bibfield  {title} {\enquote {\bibinfo {title} {{Effect of ionic
  concentration on electrokinetic instability in a cross-shaped
  microchannel}},}\ }\href {\doibase 10.1007/s10404-008-0316-2} {\bibfield
  {journal} {\bibinfo  {journal} {Microfluidics and Nanofluidics}\ }\textbf
  {\bibinfo {volume} {6}},\ \bibinfo {pages} {189--202} (\bibinfo {year}
  {2009})}\BibitemShut {NoStop}%
\bibitem [{\citenamefont {Song}\ \emph {et~al.}(2017)\citenamefont {Song},
  \citenamefont {Yu}, \citenamefont {Zhou}, \citenamefont {Antao},
  \citenamefont {Prabhakaran},\ and\ \citenamefont {Xuan}}]{Song2017}%
  \BibitemOpen
  \bibfield  {author} {\bibinfo {author} {\bibfnamefont {L.}~\bibnamefont
  {Song}}, \bibinfo {author} {\bibfnamefont {L.}~\bibnamefont {Yu}}, \bibinfo
  {author} {\bibfnamefont {Y.}~\bibnamefont {Zhou}}, \bibinfo {author}
  {\bibfnamefont {A.~R.}\ \bibnamefont {Antao}}, \bibinfo {author}
  {\bibfnamefont {R.~A.}\ \bibnamefont {Prabhakaran}}, \ and\ \bibinfo {author}
  {\bibfnamefont {X.}~\bibnamefont {Xuan}},\ }\bibfield  {title} {\enquote
  {\bibinfo {title} {{Electrokinetic instability in microchannel
  ferrofluid/water co-flows}},}\ }\href {\doibase 10.1038/srep46510} {\bibfield
   {journal} {\bibinfo  {journal} {Scientific Reports}\ }\textbf {\bibinfo
  {volume} {7}},\ \bibinfo {pages} {1--9} (\bibinfo {year} {2017})}\BibitemShut
  {NoStop}%
\bibitem [{\citenamefont {Winjet}\ \emph {et~al.}(2008)\citenamefont {Winjet},
  \citenamefont {Kaofeng}, \citenamefont {Shih},\ and\ \citenamefont
  {Yu}}]{Winjet2008}%
  \BibitemOpen
  \bibfield  {author} {\bibinfo {author} {\bibfnamefont {L.}~\bibnamefont
  {Winjet}}, \bibinfo {author} {\bibfnamefont {Y.}~\bibnamefont {Kaofeng}},
  \bibinfo {author} {\bibfnamefont {M.~H.}\ \bibnamefont {Shih}}, \ and\
  \bibinfo {author} {\bibfnamefont {K.~C.}\ \bibnamefont {Yu}},\ }\bibfield
  {title} {\enquote {\bibinfo {title} {{Microfluidic mixing utilizing
  electrokinetic instability stirred by electric potential perturbations in a
  glass microchannel}},}\ }\href@noop {} {\bibfield  {journal} {\bibinfo
  {journal} {Optoelectronics and Advanced Materials, Rapid Communications}\
  }\textbf {\bibinfo {volume} {2}},\ \bibinfo {pages} {117--125} (\bibinfo
  {year} {2008})}\BibitemShut {NoStop}%
\end{thebibliography}%

\end{document}